\documentclass[aps,pra,twocolumn,amsmath,amssymb,showpacs,floatfix]{revtex4}
\usepackage{graphicx,bm}
\renewcommand{\vec}[1]{\bm#1}
\newcommand{\abs}[1]{\lvert#1\rvert}
\newcommand{\mat}[1]{\underline{#1}}
\newcommand{\magn}{\mathcal M}
\DeclareMathOperator{\Ei}{Ei}
\DeclareMathOperator{\Li}{Li}
\DeclareMathOperator{\Tr}{Tr}
\let\Re\relax\DeclareMathOperator{\Re}{Re}

\begin{document}

\title{Transverse spin diffusion in strongly interacting Fermi gases}
\author{Tilman Enss}

\affiliation{Institut f\"ur Theoretische Physik, Universit\"at
  Heidelberg, D-69120 Heidelberg, Germany} 

\begin{abstract}
  We compute spin diffusion in a dilute Fermi gas at arbitrary
  temperature, polarization, and strong interaction in the normal phase
  using kinetic theory.  While the longitudinal spin diffusivity
  $D_\parallel$ depends weakly on polarization and diverges for small
  temperatures, the transverse spin diffusivity $D_\perp$ has a strong
  polarization dependence and approaches a finite value for $T\to0$ in
  the Fermi liquid phase.  For a 3D unitary Fermi gas at infinite
  scattering length, the diffusivities reach a minimum near the quantum
  limit of diffusion $\hbar/m$ in the quantum degenerate regime and
  are strongly suppressed by medium scattering, and we discuss the
  importance of the spin-rotation effect.  In two dimensions,
  $D_\perp$ attains a minimum at strong coupling $-1 \lesssim
  \ln(k_Fa_{2D}) \lesssim 1$ and reaches $D_\perp \sim 0.2\dotsc
  0.3\,\hbar/m$ at large polarization.  These values are consistent
  with recent measurements of two-dimensional ultracold atomic gases
  in the strong coupling regime.
\end{abstract}


\maketitle

%%%%%%%%%%%%%%%%%%%%%%%%%%%%%%%%%%%%%%%%%%%%%%%%%%%%%%%%%%%%%%%%%%%%%%%%

\section{Introduction}
\label{sec:intro}

Spin diffusion is one of the basic transport processes which tends to
compensate an imbalance of magnetization between regions of a sample.
It has been studied, e.g., in liquid helium \cite{leggett1970},
spintronics \cite{awschalom2002}, and recently in ultracold atomic
gases \cite{sommer2011a, koschorreck2013}.  If one writes the local
magnetization vector as $\vec\magn=\magn\hat{\vec e}$, the
magnetization gradient $\nabla \vec\magn = (\nabla\magn)\hat{\vec e} +
\magn\nabla \hat{\vec e}$ has two contributions: longitudinal
diffusion acts between regions of different magnitude of magnetization
$\magn$, i.e., different polarization.  Second, transverse spin diffusion
arises for spins of the same magnitude $\magn$ but different orientation
$\hat{\vec e}$, and determines the damping of transverse spin waves.
The diffusivities associated with both channels have equal magnitude
at high temperatures in the nondegenerate regime (Boltzmann limit), as
well as for an unpolarized gas.  However, they differ for the most
interesting case of a polarized gas at low temperature in the quantum
degenerate regime, since different scattering processes are
responsible for the two channels.  While the longitudinal spin
diffusivity $D_\parallel$ grows as $T^{-2}$ for a Fermi liquid at low
temperature $T$ due to Pauli blocking, the transverse spin diffusivity
$D_\perp$ is much lower---corresponding to larger spin drag---and
reaches a constant value as $T\to0$ in the normal phase, i.e., in the
absence of a phase transition.

Experiments in dilute solutions of $^3$He in liquid $^4$He can be
understood essentially within kinetic theory and the Born
approximation for weakly interacting quasiparticles.  Kinetic
equations for transverse spin transport were derived by Landau and
Silin \cite{landau1956} and applied to degenerate and/or polarized
gases \cite{leggett1970, lhuillier1982, meyerovich1985, jeon1987,
  jeon1988, jeon1989, ruckenstein1989, mullin1992, meyerovich1992,
  mullin2006}.  Transverse diffusion is influenced by the
spin-rotation effect by which the spin current precesses around the
molecular field of a polarized gas \cite{leggett1970}; a similar
effect of identical particle spin rotation occurs when two scattering
spins rotate around the common axis given by the sum of the two spins
\cite{lhuillier1982}.

Still, in dilute solutions of $^3$He strong magnetic fields are
required to reach a fully polarized state.  The advent of ultracold
atomic gases \cite{bloch2008} provides new experimental opportunities:
one can selectively drive radiofrequency transitions between atomic
hyperfine levels and coherently control the population of different
``spin'' states.  In this way, both longitudinal \cite{sommer2011a}
and transverse \cite{koschorreck2013} spin transport have recently
been measured.

Crucially, in ultracold atomic gases the scattering length can be
tuned to become much larger than the particle spacing.  In such
strongly interacting Fermi gases new transport phenomena arise, for
instance, almost perfect fluidity \cite{schaefer2009, enss2011,
  wlazlowski2012}, where the ratio of shear viscosity to entropy
density $\eta/s \gtrsim \hbar/k_B$ is bounded from below by quantum
mechanics.  The related question of whether quantum mechanics provides a
bound $D\gtrsim \hbar/m$ for spin diffusion has recently been studied
in the normal Fermi liquid phase for longitudinal \cite{duine2010,
  sommer2011a, bruun2011, bruun2012, enss2012visc, enss2012spin,
  wlazlowski2013} and transverse \cite{koschorreck2013} spin
diffusion; a sum rule for the spin conductivity is derived in
\cite{enss2013sum}.  In current experiments interactions become as
strong as allowed by quantum mechanical unitarity, and the Born
approximation is not applicable.  In this work we develop a kinetic
theory based on the \emph{many-body} $T$-matrix, building on previous
works using the $T$-matrix \cite{ruckenstein1989, mullin1992,
  meyerovich1992}, and we find a substantial suppression of the
diffusivity by medium scattering beyond the Born approximation.  The
values we obtain for the transverse spin diffusivity $D_\perp$ are
consistent with the recent spin-echo measurements of a two-dimensional
Fermi gas at strong interaction \cite{koschorreck2013}.

This paper is organized as follows: In Sec.\ \ref{sec:model} we
introduce the model of strongly interacting fermions and their
scattering in the $T$-matrix approximation, while Sec.\ \ref{sec:kin}
explains the derivation of kinetic theory for transverse and
longitudinal spin diffusion.  In Sec.\ \ref{sec:results} we present
and discuss our results and conclude in Sec.\ \ref{sec:concl}.

%%%%%%%%%%%%%%%%%%%%%%%%%%%%%%%%%%%%%%%%%%%%%%%%%%%%%%%%%%%%%%%%%%%%%%%%

\section{Model and T-matrix}
\label{sec:model}

We consider a two-component Fermi gas with contact interactions
described by the grand canonical Hamiltonian,
\begin{align}
  \label{eq:ham}
  \mathcal H
  = \sum_{\vec k\sigma} (\varepsilon_{\vec k}-\mu_\sigma)
  c_{\vec k\sigma}^\dagger c_{\vec k\sigma}
  + \frac{g_0}{V} \sum_{\vec k\vec k'\vec q}
  c_{\vec k+}^\dagger c_{\vec k'-}^\dagger
  c_{\vec k'-\vec q,-} c_{\vec k+\vec q,+}
\end{align}
with the free-particle dispersion relation $\varepsilon_{\vec k}=\vec
k^2/2m$ for particles of mass $m$.  We work in units where $\hbar = 1
= k_B$.  In a polarized gas the spin species $\sigma=\pm 1$ have
different chemical potentials $\mu_\sigma$, and we define the
effective magnetic field $h=(\mu_+ - \mu_-)/2$ conjugate to the spin
imbalance.  Motivated by experiments with ultracold atomic gases, we
consider only $s$-wave scattering, which acts between different spin
species by the Pauli principle.  The contact interaction $g_0$ needs
to be regularized in the ultraviolet both in two and three dimensions,
which is done using the $T$-matrix.

%%%%%%%%%%%%%%%%%%%%%%%%%%%%%%%%%%%%%%%%%%%%%%%%%%%%%%%%%%%%%%%%%%%%%%%% 

\subsection{Scattering cross sections}
\label{sec:scatt}

In three dimensions (3D) the vacuum, or two-body $T$-matrix, reads
\begin{align}
  \label{eq:t3d}
  \mathcal T_0(E) = \frac{4\pi/m}{a^{-1} - \sqrt{-mE}}
\end{align}
in terms of the $s$-wave scattering length $a$.  In the center-of-mass
frame, the kinetic energy of two particles with momenta $\vec k$ and
$-\vec k$ is $\omega=2\varepsilon_{\vec k}$ and $\mathcal
T_0(\omega+i0) = (4\pi/m)/(a^{-1}+ik)$ is proportional to the Landau
scattering amplitude $f(k)=-1/(a^{-1}+ik)$.  The differential cross
section in vacuum
\begin{align}
  \label{eq:dsigma3dvac}
  \frac{d\sigma}{d\Omega} = \frac{1}{a^{-2}+k^2}
\end{align}
reaches a finite value $a^2$ at low energy $k\to0$, or diverges as
$k^{-2}$ at unitarity $a^{-1}=0$.  At finite density two particles
scatter in the presence of a medium which blocks scattering into
intermediate states that are already occupied (Pauli blocking), and
one has to use the many-body $T$-matrix $\mathcal T(\vec q,\omega)$
for total momentum $\vec q$ and frequency $\omega$.  While the exact
$T$-matrix for our model \eqref{eq:ham} is not known, at sufficiently
high temperatures or in a $1/N$ expansion (see below) it is very well
approximated by summing up the particle-particle ladder diagrams
\cite{nozieres1985},
\begin{multline}
  \label{eq:tmat}
  \mathcal T^{-1}(\vec q,\omega)
  = \mathcal T_0^{-1}(E=\omega+\mu_++\mu_--\varepsilon_q/2+i0)\\
  + \int \frac{d^dk}{(2\pi)^d}
  \frac{n_{\vec k,+} + n_{\vec k+\vec q,-}}
  {\omega+\mu_++\mu_- - \varepsilon_{\vec k} - \varepsilon_{\vec k+\vec q}+i0}
\end{multline}
where $n_{\vec k\sigma} =[\exp(\beta (\varepsilon_{\vec k}-
\mu_\sigma))+1]^{-1}$ is the Fermi distribution.  In the general case,
the scattering cross section is given in terms of the many-body
$T$-matrix as
\begin{align}
  \label{eq:dsigma3d}
  \frac{d\sigma}{d\Omega} = \frac{m^2}{(4\pi)^2} 
  \abs{\mathcal T(\vec q,\omega)}^2
\end{align}
where the kinetic energy is $\omega = \varepsilon_{p_1}+
\varepsilon_{p_2}-\mu_+-\mu_- = \varepsilon_q/2+
2\varepsilon_k-\mu_+-\mu_-$ for incoming particles with momenta $\vec
p_{1,2} = \vec q/2\pm\vec k$.

In two dimensions (2D) the vacuum $T$-matrix is \cite{adhikari1986}
\begin{align}
  \label{eq:t2d}
  \mathcal T_0(E) = \frac{4\pi/m}{\ln(\varepsilon_B/E)+i\pi}
\end{align}
where $\varepsilon_B \equiv \hbar^2/ma_{2D}^2$ is the binding
energy of the two-body bound state.  In experiments a quasi-2D
geometry is realized by a strong confinement of the three-dimensional
system in one direction; well below the confinement energy,
$\varepsilon_B$ is replaced by the exact quasi-2D binding energy, which
is given in terms of the 3D scattering length $a$ and the confinement
length \cite{petrov2001}.  The $T$-matrix is related to the 2D
scattering amplitude in vacuum as $f(k) = m \mathcal
T_0(2\varepsilon_k+i0) = 4\pi/[\ln(1/k^2a_{2D}^2)+i\pi]$, and the
corresponding differential cross section is
\begin{align}
  \label{eq:dsigma2dvac}
  \frac{d\sigma}{d\Omega}
  = \frac{2\pi}{k} \, \frac{1}{\ln^2(k^2a_{2D}^2)+\pi^2}.
\end{align}
In the general case of the 2D many-body $T$-matrix \eqref{eq:tmat},
\begin{align}
  \label{eq:dsigma2d}
  \frac{d\sigma}{d\Omega}
  = \frac{m^2}{8\pi k} \abs{\mathcal T(\vec q,\omega)}^2.
\end{align}
In both 2D and 3D, the scattering cross section does not depend on the
orientation of outgoing momenta $\vec p_{3,4}=\vec q/2\pm\vec k'$;
this simplifies the angular averages in the collision integral and
precludes lateral spin rotation.

In the Boltzmann limit far above the Fermi temperature $T_F$, the
medium corrections are small and one may use the vacuum $T$-matrix.
In the quantum degenerate regime, however, medium effects become
large, and the system undergoes a phase transition toward $s$-wave
superfluidity at $T_c \simeq 0.16\,T_F$ in the 3D unpolarized unitary
Fermi gas \cite{ku2012}.  In order to include strong coupling effects
systematically in a diagrammatic approach, one option is to use a
$1/N$ expansion in the number of fermion flavors $N$ to compute the
thermodynamics above and below $T_c$ \cite{nikolic2007} as well as
transport \cite{enss2012crit}.  Eventually, the results are
extrapolated to the physical case $N=1$.  For large $N$ scattering is
weak even at unitarity and it is justified to compute transport
properties using kinetic theory consistently up to a certain order in
$1/N$; for obtaining transport coefficients to leading order one
should use the many-body $T$-matrix in the collision integral but the
free Fermi gas for the thermodynamic quantities (pressure, density,
susceptibility) that appear in transport \cite{enss2012crit}.
Interaction or mean-field corrections to the quasiparticle dispersion
relation as well as to the thermodynamic properties
\cite{nozieres1985} appear only at subleading order in the $1/N$
expansion and are therefore neglected in this work; their importance
is discussed, for instance, in Ref.\ \cite{chiacchiera2009}.

%%%%%%%%%%%%%%%%%%%%%%%%%%%%%%%%%%%%%%%%%%%%%%%%%%%%%%%%%%%%%%%%%%%%%%%% 

\subsection{Thermodynamics}
\label{sec:thermo}

We perform the transport calculation in a grand canonical setting in
terms of the dimensionless chemical potentials $\beta\mu_\pm$ and
interaction parameter $\beta\varepsilon_B$, where $\beta=1/k_BT$.  In
order to compare our results with experiments for a fixed reduced
temperature $T/T_F$, magnetic field $h/E_F$, and interaction parameter
$k_Fa$, one needs to know the equation of state
$n(\beta\mu_+,\beta\mu_-, \beta\varepsilon_B)$.  For the unpolarized
unitary Fermi gas in 3D this has been measured recently \cite{ku2012},
but it is not available with comparable accuracy for the polarized
gas.  We therefore substitute the equation of state of the free Fermi
gas, which is readily available and consistent with a $1/N$ expansion.
Indeed, at large polarization close to the polaron limit
\cite{lobo2006, schmidt2011} where the diffusivity has the most
interesting behavior, the majority species behaves almost as a free
Fermi gas, and possible phase transitions are shifted to temperatures
below the experimentally accessible range ($T \gtrsim 0.1\,T_F$).

The chemical potentials $\mu_\sigma$ for species $\sigma$ determine
the fugacities $z_\sigma = \exp(\beta\mu_\sigma)$, and hence the
pressure $P_\sigma$, density $n_\sigma$, and susceptibility
$\chi_\sigma$ of the free Fermi gas:
\begin{align}
  \label{eq:P}
  P_\sigma & =
  -\Li_{d/2+1}(-z_\sigma) \beta^{-1} \lambda_T^{-d}  \\
  \label{eq:n}
  n_\sigma & = -\Li_{d/2}(-z_\sigma) \lambda_T^{-d} \\
  \label{eq:chi}
  \chi_\sigma & = -\Li_{d/2-1}(-z_\sigma) \beta \lambda_T^{-d}
\end{align}
in terms of the thermal length $\lambda_T = \sqrt{2\pi\beta/m}$ and
the polylogarithm $\Li_s(z)$.  The (kinetic) energy density
$\varepsilon_\sigma = \int d^dp/(2\pi)^d\, \varepsilon_p n_{p\sigma} =
(d/2) P_\sigma$ by scale invariance for the free Fermi gas.  The total
density $n = n_++n_-$ and magnetization $\magn=n_+-n_-$ determine the
polarization $M=\magn/n$.  The characteristic degeneracy temperature
is the Fermi temperature $T_F = k_F^2/2m$ associated with the total
density of both spin species, $n=k_F^3/3\pi^2$ (3D) and $n=k_F^2/2\pi$
(2D), respectively.

For a typical experimental setup where the reduced temperature $T/T_F$
and the polarization $M$ are given, we first compute the total density
as
\begin{align}
  \label{eq:5}
  n\lambda_T^3 & = \frac{8}{3\sqrt\pi} (T/T_F)^{-3/2} & &
  \text{(3D)} \\
  n\lambda_T^2 & = 2(T/T_F)^{-1} & & \text{(2D)}
\end{align}
and then the component densities $n_\pm = (1\pm M)n/2$.  Inverting
Eq.~\eqref{eq:n} gives the chemical potentials $\mu_\pm$ which are
the starting point for the grand canonical calculation.  In two
dimensions, $z_\pm = \exp[(1\pm M)/(T/T_F)]-1$.

%%%%%%%%%%%%%%%%%%%%%%%%%%%%%%%%%%%%%%%%%%%%%%%%%%%%%%%%%%%%%%%%%%%%%%%%

\section{Kinetic theory}
\label{sec:kin}

The kinetic equation for particles with internal states can be written
as a matrix equation for the occupation number matrix $\mat n_p$ in
internal space.  In the case of spin-$1/2$ fermions, $\mat n_p$ is a
$2\times2$ matrix which satisfies the kinetic equation \cite{jeon1989}
\begin{multline}
  \label{eq:kin}
  \frac{D\mat n_p}{Dt}
  \equiv \frac{\partial\mat n_p}{\partial t}
  + \frac 12 [\nabla_p\mat\varepsilon_p, \nabla_r\mat n_p]_+ 
  - \frac 12 [\nabla_r\mat\varepsilon_p, \nabla_p\mat n_p]_+ \\
  + \frac{i}{\hbar} [\mat\varepsilon_p, \mat n_p]_- 
  = \left( \frac{\partial\mat n_p}{\partial t} \right)_\text{coll}.
\end{multline}
The left-hand side is the drift term, where the energy matrix 
\begin{align}
  \label{eq:dispersion}
  \mat\varepsilon_p = \varepsilon_p\mat I + \vec h_p\cdot \mat{\vec\sigma} 
\end{align}
is given in terms of the bare dispersion relation $\varepsilon_p$ and
$\vec h_p=-\tfrac{1}{2}\hbar\vec\Omega$, where $\vec\Omega=
\vec\Omega_0 +\vec\Omega_\text{mf}$ is the effective Larmor frequency
and $\mat{\vec\sigma}$ are the Pauli matrices.  The bare Larmor
frequency is $\vec\Omega_0=\gamma \vec B$ in an external magnetic
field $\vec B$, and $\vec\Omega_\text{mf}$ is the Larmor frequency due
to the molecular field of surrounding spins.  The drift term resembles
that of the Landau-Silin equation \cite{landau1956}, where the
anticommutators $[,]_+$ also include mean-field terms and the
commutator $[,]_-$ is responsible for the spin-rotation effect of
spins precessing about the effective magnetic field.  To leading order
in a $1/N$ expansion we may neglect the mean-field corrections in the
anticommutators because they are small compared to the bare dispersion
$\mat\varepsilon_p$, but the mean-field term is the leading
contribution in the spin-rotation term.

The right-hand side of Eq.~\eqref{eq:kin} is the collision integral
\begin{multline}
  \label{eq:matcoll}
  \left( \frac{\partial\mat n_{p_1}}{\partial t} \right)_\text{coll}\\
  = \frac{1}{(2\pi)^{2d-1}} \int d^dp_2\, d^dp_3\, d^dp_4\,
  \abs{\mathcal T(\vec p_1+\vec p_2,\omega)}^2\\
  \times \delta(\vec p_1+\vec p_2-\vec p_3-\vec p_4)\,
  \delta(\varepsilon_{p_1} + \varepsilon_{p_2} - \varepsilon_{p_3} -
  \varepsilon_{p_4})\\
  \times 
  \frac 14 \Bigl\{ [\tilde{\mat n}_1, \tilde{\mat n}_2^\pm]_+
  \Tr(\mat n_3 \mat n_4^\pm) - [\mat n_1, \mat n_2^\pm]_+
  \Tr(\tilde{\mat n}_3 \tilde{\mat n}_4^\pm) \Bigr\}
\end{multline}
for incoming particles $(\vec p_1,+)$ and $(\vec p_2,-)$ and outgoing
particles $(\vec p_3,+)$ and $(\vec p_4,-)$.  This expression for the
collision integral is identical to Eq.~(2.31) of
Ref.~\cite{mullin1992} specialized to fermions and using the fact that
the many-body $T$-matrix \eqref{eq:tmat} in the ladder approximation
does not depend on the direction of outgoing momenta in the
center-of-mass frame.  For atomic gases at low temperatures the
$s$-wave channel becomes dominant and only scattering between $+$ and
$-$ particles occurs; consequently, the $T$-matrix $\mathcal T(\vec
p_1+\vec p_2,\omega)$ only has components for unlike spins.  This is
reflected by the trace over spin indices $\Tr(\mat n_3 \mat n_4^\pm)$,
where $\mat n_p^\pm=\Tr(\mat n_p)\mat I-\mat n_p$: in $\mat n_p^\pm$,
the diagonal $+$ and $-$ elements of $\mat n_p$ are interchanged, and
the trace runs over unlike spins 3 and 4.  Furthermore, the fermionic
states are unoccupied with probability $\tilde{\mat n}_p=\mat I-\mat
n_p$, and the notation $\mat n_1$ stands for $\mat n_{p_1}$ etc.  In
the case of longitudinal spin diffusion the collision integral becomes
diagonal in the spin indices.  However, for transverse spin diffusion
the collision integral acquires off-diagonal terms and the full
occupation matrix $\mat n_p$ needs to be kept.

One may parametrize the occupation matrix $\mat n_p$ in terms of
particle $f_p$ and spin $\vec\sigma_p$ variables
\begin{align}
  \label{eq:nparam}
  \mat n_p = \frac 12 \left( f_p \mat I + \vec\sigma_p \cdot
    \mat{\vec\sigma} \right),
\end{align}
and the kinetic equation \eqref{eq:kin} may be written in components
\begin{multline}
  \label{eq:driftf}
  \frac{Df_p}{Dt} \equiv \frac{\partial f_p}{\partial t} + \sum_i
  \left[ \frac{\partial\varepsilon_p}{\partial p_i} \frac{\partial
      f_p}{\partial r_i} - \frac{\partial\varepsilon_p}{\partial r_i}
    \frac{\partial f_p}{\partial p_i} \right. \\
  \left. + \frac{\partial\vec
      h_p}{\partial p_i} \cdot \frac{\partial\vec\sigma_p}{\partial
      r_i} - \frac{\partial\vec h_p}{\partial r_i} \cdot
    \frac{\partial\vec\sigma_p}{\partial p_i} \right]
  = \left( \frac{\partial f_p}{\partial t} \right)_\text{coll}
\end{multline}
and
\begin{multline}
  \label{eq:driftsigma}
  \frac{D\vec\sigma_p}{Dt}
  \equiv \frac{\partial\vec\sigma_p}{\partial t}
  + \sum_i \left[ \frac{\partial\varepsilon_p}{\partial p_i}
    \frac{\partial\vec\sigma_p}{\partial r_i}
    - \frac{\partial\varepsilon_p}{\partial r_i}
    \frac{\partial\vec\sigma_p}{\partial p_i} \right. \\
  \left. + \frac{\partial \vec h_p}{\partial p_i} 
    \frac{\partial f_p}{\partial r_i}
    - \frac{\partial \vec h_p}{\partial r_i} 
    \frac{\partial f_p}{\partial p_i} \right] 
  - \frac{2}{\hbar} \vec h_p \times \vec\sigma_p 
  = \left( \frac{\partial\vec\sigma_p}{\partial t} \right)_\text{coll}.
\end{multline}
The local magnetization is $\vec\magn(\vec r,t) =\int
d^dp\,\vec\sigma_p/(2\pi)^d =\magn(\vec r,t)\hat{\vec e}(\vec r,t)$
and we choose the local magnetization direction $\hat{\vec e}(\vec
r,t)$ as the spin quantization axis, such that the local equilibrium
distribution matrix $\mat n_p^0$ is diagonal with entries $n_{p+}$ and
$n_{p-}$.  Note that $\vec\magn$ need not be parallel to an external
magnetic field $\vec B$.  According to Eq.~\eqref{eq:nparam}, $f_p^0
= n_{p+} + n_{p-}$ and $\vec\sigma_p^0 = (n_{p+} - n_{p-}) \hat{\vec
  e}$.  The gradient of the magnetization has two contributions, the
longitudinal and transverse parts
\begin{align}
  \label{eq:gradm}
  \frac{\partial \vec\magn}{\partial r_i} = \frac{\partial \magn}{\partial
    r_i} \hat{\vec e} + \magn \frac{\partial \hat{\vec e}}{\partial r_i}.
\end{align}
We linearize the kinetic equations \eqref{eq:driftf} and
\eqref{eq:driftsigma} around the local equilibrium distribution, $\mat
n_p=\mat n_p^0+\delta \mat n_p$, and write the drift terms as
\begin{align}
  \label{eq:driftf2}
  \frac{Df_p}{Dt}
  \equiv \frac{\partial f_p}{\partial t}
  - \sum_i v_{pi} \frac{\partial\magn}{\partial r_i} \sum_\sigma 
  \sigma t_\sigma \frac{\partial n_{p\sigma}}{\partial \varepsilon_p}
  = \left( \frac{\partial f_p}{\partial t} \right)_\text{coll}
\end{align}
and
\begin{multline}
  \label{eq:driftsigma2}
  \frac{D\vec\sigma_p}{Dt}
  \equiv \frac{\partial\vec\sigma_p}{\partial t}
  - \sum_i v_{pi} \frac{\partial\magn}{\partial r_i} \hat{\vec e} 
  \sum_\sigma t_\sigma \frac{\partial
    n_{p\sigma}}{\partial\varepsilon_p} \\
  + \sum_i v_{pi} \frac{\partial \hat{\vec e}}{\partial r_i} 
  (n_{p+} - n_{p-})
  + \vec\Omega \times \vec\sigma_p 
  = \left( \frac{\partial\vec\sigma_p}{\partial t} \right)_\text{coll}
\end{multline}
up to corrections of order $\mathcal O(\delta \mat n_p)$.  The second
(longitudinal) and third (transverse) terms in
Eq.~\eqref{eq:driftsigma2} result from the gradient of the local
magnetization \eqref{eq:gradm}.  The derivative $\partial
n_{p\sigma}/\partial \varepsilon_p$ in the longitudinal term restricts
the momentum integrals in the degenerate regime to a neighborhood of
the Fermi surface.  In contrast, in the transverse term
$n_{p+}-n_{p-}$ is nonzero everywhere between the majority and
minority Fermi surfaces, hence the phase space for scattering at low
temperature and the transverse scattering rate $\tau_\perp^{-1}$ are
larger than in the longitudinal case \cite{meyerovich1985}.

In the derivation we have used the Gibbs-Duhem relation $\sum_\sigma
n_\sigma (\partial\mu_\sigma/\partial r_i) = 0$ and
\begin{align}
  \frac{\partial n_\sigma}{\partial r_i}
  & = \chi_\sigma \frac{\partial\mu_\sigma}{\partial r_i}, &
  \chi_\sigma & = \frac{\partial n_\sigma}{\partial \mu_\sigma}, \\
  \frac{\partial \mu_\sigma}{\partial r_i}
  & = \sigma t_\sigma \frac{\partial\magn}{\partial r_i}, &
  t_\sigma & = \frac{1/n_\sigma}{\chi_+/n_++\chi_-/n_-}.
\end{align}
It then follows that
\begin{align}
  \label{eq:vel}
  \frac{\partial\varepsilon_p}{\partial p_i} & = \frac{p_i}{m} =
  v_{pi} \\
  \frac{\partial f_p^0}{\partial r_i} & = -\sum_\sigma \frac{\partial
    n_{p\sigma}}{\partial \varepsilon_p} \frac{\partial
    \mu_\sigma}{\partial r_i} = -\frac{\partial\magn}{\partial r_i}
  \sum_\sigma \sigma t_\sigma
  \frac{\partial n_{p\sigma}}{\partial \varepsilon_p} \\
  \frac{\partial \vec\sigma_p^0}{\partial r_i} & = \frac{\partial
    (n_{p+} - n_{p-})}{\partial r_i} \hat{\vec e}
  + (n_{p+} - n_{p-}) \frac{\partial
    \hat{\vec e}}{\partial r_i} \notag \\
  & = -\frac{\partial\magn}{\partial r_i}
  \hat{\vec e} \sum_\sigma t_\sigma \frac{\partial
    n_{p\sigma}}{\partial \varepsilon_p} + \frac{\partial \hat{\vec
      e}}{\partial r_i} (n_{p+} - n_{p-})
\end{align}
and we have assumed a constant $\vec h_p$.

The particle and spin currents are defined as the velocity weighted by
the distribution functions,
\begin{align}
  \label{eq:jptcl}
  J_j & = \int \frac{d^dp}{(2\pi)^d}\, v_{pj} f_p \\
  \label{eq:jspin}
  \vec J_j
  & = \int \frac{d^dp}{(2\pi)^d}\, v_{pj} \vec\sigma_p
\end{align}
for a magnetization gradient in direction $j=x,y,z$.  We shall not
consider the particle current further and instead concentrate on the
spin current.  The continuity equation for the spin density
(magnetization) is
\begin{align}
  \label{eq:contin}
  \frac{\partial\vec\magn}{\partial t} + \sum_j \frac{\partial\vec
    J_j}{\partial r_j} + \vec\Omega_0 \times \vec\magn = 0.
\end{align}
The momentum integral over the Boltzmann equation
\eqref{eq:driftsigma2} weighted by the velocity $v_{pj}$ yields the
time evolution of the spin current,
\begin{multline}
  \label{eq:boltzmannJ}
  \frac{D\vec J_j}{Dt}
  \equiv \frac{\partial\vec J_j}{\partial t}
  + \alpha_\parallel \frac{\partial\magn}{\partial r_j} \hat{\vec e}
  + \alpha_\perp \magn \frac{\partial \hat{\vec e}}{\partial r_j}\\
  + (\vec\Omega_0+\vec\Omega_\text{mf}) \times \vec J_j
  = \int \frac{d^dp}{(2\pi)^d}\, v_{pj} \left( \frac{\partial
      \vec\sigma_p}{\partial t} \right)_\text{coll}
\end{multline}
with coefficients
\begin{align}
  \label{eq:alphapara}
  \alpha_\parallel & = \int \frac{d^dp}{(2\pi)^d} \sum_i v_{pi} v_{pj}
  \sum_\sigma t_\sigma \frac{\partial
    n_{p\sigma}}{\partial\varepsilon_p}
  = \frac{2/m}{\chi_+/n_+ + \chi_-/n_-}\\
  \label{eq:alphaperp}
  \alpha_\perp & = \frac{1}{\magn} \int \frac{d^dp}{(2\pi)^d}
  \sum_i v_{pi} v_{pj} (n_{p+}-n_{p-}) 
  = \frac{P_+ - P_-}{m\magn}
\end{align}
for a free Fermi gas.  Both $\alpha_\parallel$ and $\alpha_\perp$
approach $1/m\beta$ in the Boltzmann limit and $n/m\chi$ for the
unpolarized gas.

The collision integral on the right-hand side of
Eq.~\eqref{eq:boltzmannJ} determines how the spin current relaxes by
collisions, and one has to parametrize the decay by separate time
constants $\tau_\parallel$ and $\tau_\perp$ for longitudinal and
transverse relaxation \cite{jeon1988},
\begin{align}
  \label{eq:sigmacoll}
  \int \frac{d^dp}{(2\pi)^d}\, v_{pj} 
  \left( \frac{\partial\vec\sigma_p}{\partial t} \right)_\text{coll} 
  = -\frac{1}{\tau_\parallel}(\vec J_j\cdot \hat{\vec e})\hat{\vec e}
  -\frac{1}{\tau_\perp}(\vec J_j\cdot \hat{\vec g}_j)\hat{\vec g}_j.
\end{align}
The unit vector 
\begin{align}
  \label{eq:gvector}
  \hat{\vec g}_j = x\frac{\partial \hat{\vec e}}{\partial r_j}
  + y\hat{\vec e} \times \frac{\partial \hat{\vec e}}{\partial r_j}
\end{align}
lies in the plane perpendicular to the local magnetization direction
$\hat{\vec e}$, at an angle determined by the coefficients $x$ and $y$.

In order to solve Eq.~\eqref{eq:boltzmannJ}, consider first the
rotation term $(\vec\Omega_0 + \vec\Omega_\text{mf}) \times \vec J_j$
where the molecular field $\vec\Omega_\text{mf}
=\Omega_\text{mf}\,\hat{\vec e}$ is parallel to the local
magnetization $\vec\magn$.  Hence, $\vec\magn$ in
Eq.~\eqref{eq:contin} precesses only about the external magnetic
field $\vec\Omega_0$ but not about $\vec\Omega_\text{mf}$.  In
contrast, the spin current $\vec J_j$ is in general not parallel to
$\vec\magn$ and can precess also about the molecular field
$\vec\Omega_\text{mf}$.  It is convenient to work in a frame rotating
with the external field $\vec\Omega_0$ in spin space such that the
time evolution of $\vec\magn$ approaches a quasi steady state
\cite{leggett1970}.  In the same rotating frame, $\partial \vec
J_j/\partial t = -\vec\Omega_0 \times \vec J_j$ cancels the free
precession of $\vec J_j$ in Eq.~\eqref{eq:boltzmannJ}, but the spin
current still precesses about $\vec\Omega_\text{mf}$.  This causes the
spin-rotation effect in transverse diffusion, in contrast to
longitudinal diffusion where $\vec J_j \parallel \vec\magn$ and spin
rotation is absent.  Via the continuity equation \eqref{eq:contin} for
the spin density, spin rotation in $\vec J_j$ causes a similar effect
in $\vec\magn$.  Equations \eqref{eq:boltzmannJ} and
\eqref{eq:sigmacoll} are solved by the spin current \cite{jeon1988,
  mullin1992}
\begin{align}
  \label{eq:jspinsol}
  \vec J_j
  = -D_\parallel \frac{\partial\magn}{\partial r_j} \hat{\vec e}
  - \frac{D_\perp^0}{1+\mu^2}\, \magn \left[ \frac{\partial \hat{\vec
        e}}{\partial r_j} + \mu \hat{\vec e} \times \frac{\partial
      \hat{\vec e}}{\partial r_j} \right]
\end{align}
with diffusion coefficients $D_\parallel = \alpha_\parallel
\tau_\parallel$ and $D_\perp^0 = \alpha_\perp \tau_\perp$.  The full
transverse diffusion coefficient, including the spin-rotation effect, is
given by
\begin{align}
  \label{eq:fulldiff}
  D_\perp = \frac{D_\perp^0}{1+\mu^2}
\end{align}
where the spin-rotation parameter
\begin{align}
  \mu = -\Omega_\text{mf}\, \tau_\perp
\end{align}
determines how the spin current is rotated in the plane perpendicular
to the local magnetization.  (This parameter is denoted as $\mu M$ in
other works \cite{leggett1970, jeon1988, mullin1992}, but we have
included the polarization $M$ in the definition of $\mu$.)  An example
of how
the spin-rotation effect lowers the transverse diffusivity is shown in
Sec.\ \ref{sec:results}.  Without molecular field there is no
spin-rotation effect, $\mu=0$ and $D_\perp = \alpha_\perp \tau_\perp$.

One may parametrize the deviation from local equilibrium as
\begin{align}
  \label{eq:dnp}
  \delta\mat n_p = \frac 12 ( \delta f_p\, \mat I +
  \delta\vec\sigma_p\cdot \mat{\vec\sigma} ).
\end{align}
The deviations $\delta f_p$ and $\delta\vec\sigma_p$ should overlap
with the drift terms in Eqs.~\eqref{eq:driftf2} and
\eqref{eq:driftsigma2}, and we choose the variational trial functions
\cite{jeon1988}
\begin{align}
  \delta f_p & = c \sum_i v_{pi} \frac{\partial\magn}{\partial r_i}
  \sum_\sigma \sigma t_\sigma \frac{\partial n_{p\sigma}}{\partial
    \varepsilon_p} \notag \\
  \label{eq:dsigma}
  \delta\vec\sigma_p & = \delta\vec\sigma_p^\parallel +
  \delta\vec\sigma_p^\perp 
\end{align}
with the longitudinal part
\begin{align}
  \label{eq:dsigmalong}
  \delta\vec\sigma_p^\parallel = c_\parallel \sum_i v_{pi}
  \frac{\partial\magn}{\partial r_i} \hat{\vec e} \sum_\sigma
    t_\sigma \frac{\partial n_{p\sigma}}{\partial\varepsilon_p}
\end{align}
and transverse part
\begin{align}
  \label{eq:dsigmaperp}
  \delta\vec\sigma_p^\perp = c_\perp \sum_i v_{pi} \hat{\vec g}_i
  (n_{p+}-n_{p-}).
\end{align}
In the following we shall linearize the collision integral
\eqref{eq:matcoll} for these small deviations from the equilibrium
distribution, first in the transverse and then in the longitudinal
channel.

Let us briefly discuss the assumptions and approximations involved in
the derivation of kinetic theory: we assume $(i)$ applicability of the
general hypotheses of Fermi liquid theory and the quasiparticle
picture; this condition is met in the normal phase sufficiently far
above a possible phase transition to a low-temperature symmetry broken
phase; $(ii)$ total spin conservation; $(iii)$ hydrodynamic
conditions, \emph{i.e.}, slow variations in time and space; $(iv)$
linearization of the Boltzmann equation, \emph{i.e.}, a small
departure from the local equilibrium distribution; $(v)$ ladder
approximation for the many-body $T$-matrix \eqref{eq:tmat};
consequently, the $T$-matrix does not depend on the direction of
outgoing particles in the center-of-mass frame; $(vi)$ no mean-field
drift terms except for the spin-rotation term; $(vii)$ the variational
ansatz for the deviation from equilibrium, Eqs.~\eqref{eq:dsigmalong}
and \eqref{eq:dsigmaperp}; and $(viii)$ no off-energy shell terms in
the collision integral \cite{mullin1992}.  Both the ladder
approximation and the absence of mean-field drift terms are justified
as the leading order of a low-density expansion
\cite{ruckenstein1989}, or of a systematic $1/N$ expansion in the
number of fermion flavors \cite{enss2012crit}.  Once these assumptions
are made, the kinetic theory applies to arbitrary temperature from the
Boltzmann to the degenerate limit, arbitrary polarization, anisotropic
spin-current relaxation times $\tau_\parallel$ and $\tau_\perp$, and
arbitrary $s$-wave scattering lengths $a$ beyond the Born
approximation, as long as the quasiparticle picture remains valid.

Note that the lateral spin-rotation term in the collision
integral \cite{lhuillier1982, mullin1992} only appears if
the $T$-matrix is complex and depends on direction; it vanishes in our
case for a direction-independent $T$-matrix, just as it does for a
real effective potential \cite{mullin1992}.

%%%%%%%%%%%%%%%%%%%%%%%%%%%%%%%%%%%%%%%%%%%%%%%%%%%%%%%%%%%%%%%%%%%%%%%%

\subsection{Transverse diffusion}
\label{sec:trans}

The linearized form of the collision integral \eqref{eq:matcoll} for
the $T$-matrix \eqref{eq:tmat} differs from the Born approximation in
that only $+$ and $-$ particles can scatter,
\begin{multline}
  \label{eq:dmatcoll}
  \left( \frac{\partial \delta \mat n_{p_1}}{\partial t}
  \right)_\text{coll} \\
  = \frac{1}{(2\pi)^{2d-1}} \int d^dp_2\, d^dp_3\,
  d^dp_4\, \abs{\mathcal T(\vec p_1+\vec p_2,\omega)}^2\\
  \times \delta(\vec p_1+\vec p_2-\vec p_3-\vec p_4)\,
  \delta(\varepsilon_{p_1} + \varepsilon_{p_2} - \varepsilon_{p_3} -
  \varepsilon_{p_4})\\
  \times 
  \frac 14 \Bigl\{ [ \delta\tilde{\mat n}_2^\pm \tilde{\mat n}_1 +
  \tilde{\mat n}_2^\pm \delta\tilde{\mat n}_1 + \tilde{\mat n}_1
  \delta\tilde{\mat n}_2^\pm + \delta\tilde{\mat n}_1 \tilde{\mat n}_2^\pm ]
  \Tr( \mat n_4^\pm \mat n_3) \\
  -[ \delta\mat n_2^\pm \mat n_1 + \mat n_2^\pm \delta\mat n_1 + \mat n_1
  \delta\mat n_2^\pm + \delta\mat n_1 \mat n_2^\pm ] \Tr( \tilde{\mat n}_4^\pm
  \tilde{\mat n}_3) \Bigr\}.
\end{multline}
On the right-hand side a transverse variation of the distribution
matrix is inserted using the variational ansatz in
Eq.~\eqref{eq:dsigmaperp}:
\begin{multline}
  \label{eq:dnpperp}
  \delta \mat n_p^\perp
  = \frac 12 \delta\vec\sigma_p^\perp \cdot \mat{\vec\sigma} 
  = \frac{c_\perp}{2} (n_{p+}-n_{p-}) \sum_i v_{pi} \hat{\vec g}_i \cdot
  \mat{\vec\sigma} \\
  = (n_{p+}-n_{p-}) 
  \begin{pmatrix}
    0 & s_p^* \\
    s_p & 0
  \end{pmatrix}
\end{multline}
with $s_p = s_{px} + is_{py}$ and $\vec s_p = (c_\perp/2) (v_{px}
\hat{\vec g}_x + v_{py} \hat{\vec g}_y)$.  A typical term in the
collision integral \eqref{eq:dmatcoll} has the form \cite{jeon1988}
\begin{align}
  \label{eq:collterm}
  \delta\mat n_1 \mat n_2^\pm
  & = (n_{1+}-n_{1-})
  \begin{pmatrix}
    0 & s_1^* n_{2-} \\
    s_1 n_{2+} & 0
  \end{pmatrix}, \\
  [\delta\mat n_1, \mat n_2^\pm]_+
  & = (n_{1+}-n_{1-}) (n_{2+}+n_{2-})
  \begin{pmatrix}
    0 & s_1^* \\
    s_1 & 0
  \end{pmatrix}.
\end{align}
From $(\delta\mat n_p^\perp)^\pm = \Tr(\delta\mat n_p^\perp) \mat I -
\delta\mat n_p^\perp$ follows $\delta \vec\sigma_p^{\perp\pm} =
-\delta \vec\sigma_p^\perp$, and the matrix product in the curly
brackets in Eq.~\eqref{eq:dmatcoll} becomes
\begin{multline}
  \label{eq:transoccu}
  \frac{c_\perp}{2} \sum_i \Bigl\{ \Bigl[(\tilde n_{1+}-\tilde n_{1-})
  (\tilde n_{2+}+\tilde n_{2-}) v_{1i} \\
    - (\tilde n_{1+}+\tilde n_{1-}) (\tilde n_{2+}-\tilde n_{2-})
    v_{2i}\Bigr] (n_{3+} n_{4-} + n_{3-} n_{4+}) \\
    - \Bigl[(n_{1+}-n_{1-}) (n_{2+}+n_{2-}) v_{1i} \\
    - (n_{1+}+n_{1-}) (n_{2+}-n_{2-}) v_{2i}\Bigr] (\tilde n_{3+} \tilde
    n_{4-} + \tilde n_{3-} \tilde n_{4+})
  \Bigr\} \hat{\vec g}_i \cdot \mat{\vec\sigma}.
\end{multline}
Using $\tilde n_{1+} \tilde n_{2-} n_{3+} n_{4-} =
n_{1+} n_{2-} \tilde n_{3+} \tilde n_{4-}$ from energy conservation and
\begin{align}
  \label{eq:32}
  \frac{n_{p+} \tilde n_{p-}}{n_{p-} \tilde n_{p+}} = \exp(2\beta h)
\end{align}
one may rewrite Eq.~\eqref{eq:transoccu} as
\begin{multline}
  \label{eq:transoccu2}
  -2c_\perp \sinh(\beta h) \sum_i \bigl[ e^{-\beta h} n_{1+} n_{2+}
  + e^{\beta h} n_{1-} n_{2-}\bigr] \\
  \times \tilde n_{3+} \tilde n_{4-}
  (v_{1i} - v_{2i}) \hat{\vec g}_i \cdot \mat{\vec\sigma}.
\end{multline}
The unusual occupation factors $n_{1\pm} n_{2\pm}$ are characteristic
of transverse spin diffusion and appear even though spin is conserved
during scattering.

The collision integral determines the relaxation of the transverse
current according to Eq.~\eqref{eq:sigmacoll} with the variational
form \eqref{eq:dsigmaperp} also on the right-hand side,
\begin{align}
  \frac{D\vec J_j^\perp}{Dt}
  & = \int \frac{d^dp}{(2\pi)^d} v_{pj} \Tr\left[ \mat{\vec\sigma}
    \left( \frac{\partial \delta\mat n_p^\perp}{\partial t}
    \right)_\text{coll} \right] \notag \\
  \label{eq:DJperp}
  & = -\frac{c_\perp \alpha_\perp \magn}{\tau_\perp} \hat{\vec g}_j
  = -\frac{\vec J_j^\perp}{\tau_\perp},
\end{align}
and hence the transverse scattering rate is given by
\begin{multline}
  \label{eq:tauperpgen}
  \frac{1}{\tau_\perp}
  = \frac{\sinh(\beta h)}{\alpha_\perp \magn}
  \frac{1}{(2\pi)^{3d-1}} \int d^dp_1\dotsc d^dp_4\\
  \times \delta(\vec p_1+\vec p_2-\vec p_3-\vec p_4)\,
  \delta(\varepsilon_{p_1} + \varepsilon_{p_2} - \varepsilon_{p_3} -
  \varepsilon_{p_4}) \\
  \times \abs{\mathcal T(\vec p_1+\vec p_2,\omega)}^2\,
  [e^{-\beta h} n_{1+} n_{2+} + e^{\beta h} n_{1-} n_{2-}] \\
  \times \tilde n_{3+} \tilde n_{4-} v_{1j} (v_{1j} - v_{2j}).
\end{multline}
The integral over outgoing momenta yields
\begin{multline}
  \label{eq:tauperpinc}
  \frac{1}{\tau_\perp}
  = \frac{\sinh(\beta h)}{(2\pi)^{2d}\alpha_\perp \magn}
  \int d^dp_1\, d^dp_2\, d\Omega\, \frac{\abs{\vec p_1-\vec p_2}}{m}\,
  \frac{d\sigma}{d\Omega} \\
  \times [e^{-\beta h} n_{1+} n_{2+} + e^{\beta h} n_{1-} n_{2-}]
  \tilde n_{3+} \tilde n_{4-} v_{1j} (v_{1j} - v_{2j})
\end{multline}
or in center-of-mass coordinates $\vec p_{1,2}=\vec q/2\pm \vec k$,
$\vec p_{3,4}=\vec q/2\pm \vec k'$
\begin{multline}
  \label{eq:tauperp}
  \frac{1}{\tau_\perp} = \frac{\sinh(\beta h)}{(2\pi)^{2d}\alpha_\perp \magn} 
  \int d^dq\, d^dk\, d\Omega\,
  \frac{2k}{m}\, \frac{d\sigma}{d\Omega} \\
  \left[e^{-\beta h} n_{1+} n_{2+} + e^{\beta h} n_{1-} n_{2-}\right]
  \tilde n_{3+} \tilde n_{4-} \frac{2k_j^2}{m^2}.
\end{multline}
For $T$-matrix scattering the cross section does not depend on the
angle $\Omega$ between $\vec k$ and $\vec k'$, so one can perform the
angular integrations explicitly for the Fermi distribution
$n_{p\sigma}$ and obtain (no summation over $j$)
\begin{multline}
  \label{eq:angularperp}
  \int d\Omega_{\vec q}\, d\Omega_{\vec k}\, d\Omega\, 
  \left[e^{-\beta h} n_{1+} n_{2+} + e^{\beta h} n_{1-} n_{2-}\right]\\
  \times (1-n_{3+})(1-n_{4-}) k_j^2 \\
  = \frac{S_d^3}{d} k^2 [I_{\ell=0}(a-c,b,0)+I_{\ell=0}(a+c,b,0)]
  I_{\ell=0}(a,b,c)
\end{multline}
with $a=\beta(\varepsilon_{q/2}+\varepsilon_k-(\mu_++\mu_-)/2)$,
$b=\beta\sqrt{\varepsilon_q \varepsilon_k}$, $c=\beta h$, and solid
angle $S_d$ in $d$ dimensions.  The $\ell$-wave angular averages are
given by \cite{enss2012crit}
\begin{align}
  \label{eq:Iell3d}
  I_\ell & = \frac 14 \int_{-1}^1 dx\,
  \frac{P_\ell(x)}{\cosh(a)+\cosh(bx+c)} &
  & \text{(3D)} \\
  \label{eq:Iell2d}
  I_\ell & = \frac{1}{4\pi} \int_0^{2\pi} d\phi\,
  \frac{P_\ell(\cos\phi)}{\cosh(a)+\cosh(b\cos\phi+c)} &
  & \text{(2D)}
\end{align}
with Legendre polynomials $P_\ell(x)$.  In three dimensions these
integrals are known analytically, in particular
\begin{align}
  \label{eq:Iell0}
  I_{\ell=0}(a,b,c) = \frac{1}{4b\sinh(a)}
  \ln\frac{\cosh(a+b)+\cosh(c)} {\cosh(a-b)+\cosh(c)}
\end{align}
and analytical expressions involving polylogarithms for $\ell>0$
\cite{enss2012crit}, while in two dimensions the $I_\ell$ are readily
evaluated numerically.  This leads to the transverse scattering time
\begin{multline*}
  \frac{1}{\tau_\perp} =
  \frac{4S_d^3\sinh(\beta h)}{d(2\pi)^{2d}m^2(P_+-P_-)}
  \int_0^\infty dq\, q^{d-1} 
  \int_0^\infty dk\, k^{d+2} \\
  \times \frac{d\sigma}{d\Omega}
  [I_0(a-c,b,0)+I_0(a+c,b,0)]I_0(a,b,c)
\end{multline*}
using $\alpha_\perp$ from Eq.~\eqref{eq:alphaperp}.  Finally, the
diffusion coefficient is given by $D_\perp^0 = \alpha_\perp \tau_\perp$.

\subsubsection{Limiting cases}

The expression for the scattering rate simplifies in two limits: the
Boltzmann limit $T\gg T_F$, and the unpolarized limit $\beta h\to0$.
In the Boltzmann limit,
\begin{align}
  \label{eq:Iboltz}
  I_\ell(a,b,c) \to \delta_{\ell,0} \exp(-a)
\end{align}
and the angular average $2\sinh(\beta h)[I_0^-+I_0^+]I_0/(P_+-P_-) \to
\beta \lambda_T^{2d} n \exp(-\beta \varepsilon_q/2) \exp(-2\beta
\varepsilon_k)$ such that
\begin{multline}
  \label{eq:tauperpboltz}
  \frac{1}{\tau_\perp}
  = \frac{2S_d^3\beta\lambda_T^{2d}n}{d(2\pi)^{2d}m^2} 
  \int_0^\infty dq\, q^{d-1} \exp(-q^2\lambda_T^2/8\pi) \\
  \times \int_0^\infty dk\, k^{d+2} \exp(-k^2\lambda_T^2/2\pi)
  \frac{d\sigma}{d\Omega}.
\end{multline}
In the Boltzmann limit the medium effect on scattering becomes small
and one may use the vacuum scattering cross section
\eqref{eq:dsigma3dvac}, which depends only on the relative momentum $k$
but not on the center-of-mass momentum $q$, and the integrals are
readily performed in 3D to yield
\begin{align}
  \label{eq:tauperp3Dboltz}
  \frac{1}{\tau_\perp}
  & = \frac{2\sqrt 2 n \lambda_T^7}{3\pi^3\beta} 
  \int_0^\infty dk\, k^5
  \frac{\exp(-k^2\lambda_T^2/2\pi)}{a^{-2} + k^2} \\
  & = \frac{4\sqrt 2 n \lambda_T^3}{3\pi\beta}
  \bigl[ 1 - \beta\varepsilon_B - (\beta\varepsilon_B)^2
    \exp(\beta\varepsilon_B) \Ei(-\beta\varepsilon_B) \bigr] \notag
\end{align}
where $\Ei(x)$ is the exponential integral, and we have defined a
``binding energy'' $\varepsilon_B=\hbar^2/ma^2$ also on the BCS side
for negative $a$, where there is no two-body bound state.  For the
unitary gas $\beta\varepsilon_B=0$ and the expression in brackets is
unity, corresponding to an effective cross section $\sigma =
\lambda_T^2$, and we obtain the transverse scattering time and
diffusivity (with $\alpha_\perp=1/m\beta$):
\begin{align}
  \label{eq:tauperp3D}
  \tau_\perp & = \frac{9\pi^{3/2}\hbar}{32\sqrt2 k_BT_F}
  \left(\frac{T}{T_F}\right)^{1/2}, \\
  D_\perp^0 & = \frac{9\pi^{3/2}\hbar}{32\sqrt2 m}
  \left(\frac{T}{T_F}\right)^{3/2} & & \text{(3D)}.
\end{align}
These results coincide with the longitudinal scattering time and
diffusivity in the Boltzmann limit \cite{bruun2011, sommer2011a}.  In
the weak-coupling limit the scattering cross section is $4\pi a^2$ and
the term in parentheses in Eq.~\eqref{eq:tauperp3Dboltz} approaches
$4\pi a^2/\lambda_T^2$.

In two dimensions we find in the Boltzmann limit
\begin{align}
  \label{eq:tauperp2D}
  \frac{1}{\tau_\perp}
  & = \frac{n\lambda_T^2}{\pi\beta}
  \lambda_T^4 \int dk\, k^3\,
  \frac{\exp(-k^2\lambda_T^2/2\pi)}{\ln^2(k^2a_{2D}^2)+\pi^2} \\
  & = \frac{2\pi n\lambda_T^2}{\beta Q}
  = \frac{4\pi k_BT_F}{Q} \notag
\end{align}
with
\begin{align}
  \label{eq:17}
  Q = \ln^2(2\beta\varepsilon_B/3)+\pi^2
\end{align}
evaluated at the saddle point of the $k$ integral
\cite{schaefer2012}.  The scattering time and diffusivity
\begin{align}
  \label{eq:Dperp2D}
  \tau_\perp & = \frac{\hbar Q}{4\pi k_BT_F}, & 
  D_\perp^0 & = \frac{\hbar Q}{4\pi m} \frac{T}{T_F} &
  & \text{(2D)}
\end{align}
again agree with the longitudinal scattering time and diffusivity in
the Boltzmann limit \cite{bruun2012, enss2012visc}.

The second limit where $\tau_\perp$ simplifies is the unpolarized
limit $\beta h\to 0$ at arbitrary temperature in the normal phase
$T>T_c$.  The prefactor $\sinh(\beta h)/(P_+-P_-) \to \beta/n$, and
the angular average becomes $[I_0^-+I_0^+]I_0\to 2I_0^2(a,b,c=0)$:
\begin{align}
  \label{eq:tauperpunpol}
  \frac{1}{\tau_\perp}
  = \frac{8S_d^3\beta}{d(2\pi)^{2d}m^2n} 
  \int dq\, q^{d-1} \int dk\, k^{d+2}\,
  \frac{d\sigma}{d\Omega}\, I_0^2.
\end{align}
We shall see below in Sec.~\ref{sec:long} that this coincides with
the longitudinal scattering rate in the unpolarized limit.

\subsubsection{Spin rotation}

The transverse diffusivity $D_\perp$ is modified by the spin-rotation
effect where the spin current $\vec J_j$ precesses around the
molecular field $\vec\Omega_\text{mf} =\Omega_\text{mf}\,\hat{\vec
  e}$.  The field acting on spin $1$ due to interaction with
surrounding spins $2$ reads
\begin{align}
  \label{eq:1}
  \vec\Omega_1 = \int \frac{d^dp_2}{(2\pi)^d}
  \Re \mathcal T(\vec p_1+\vec p_2,\omega) \vec\sigma_2
\end{align}
with $\omega=\varepsilon_{p_1}+\varepsilon_{p_2}-\mu_+-\mu_-$.  The
resulting spin rotation term in the time evolution of $\vec\sigma_1$
\eqref{eq:driftsigma2} is then
\begin{multline}
  \label{eq:8}
  \frac{D\vec\sigma_1}{Dt}\Bigr\rvert_\text{spinrot}
  = \vec\Omega_1 \times \vec\sigma_1 \\
  = \int \frac{d^dp_2}{(2\pi)^d}
  \Re \mathcal T(\vec p_1+\vec p_2,\omega)
  [\vec\sigma_2 \times \vec\sigma_1].
\end{multline}
We expand $\vec\sigma_p = \vec\sigma_p^0 + \delta\vec\sigma_p^\perp$
with local equilibrium distribution $\vec\sigma_p^0 =
(n_{p+}-n_{p-})\hat{\vec e}$ and small deviation \eqref{eq:dsigmaperp}
to linear order,
\begin{multline}
  \label{eq:9}
  \vec\sigma_2 \times \vec\sigma_1
  = \vec\sigma_2^0 \times \delta\vec\sigma_1^\perp
  + \delta\vec\sigma_2^\perp \times \vec\sigma_1^0 \\
  = \sum_i (v_{1i}-v_{2i})(n_{1+}-n_{1-})(n_{2+}-n_{2-}) \hat{\vec e}
  \times \hat{\vec g}_i.
\end{multline}
The time evolution of the transverse spin current
\begin{align}
  \label{eq:4}
  \frac{D\vec J_j^\perp}{Dt}\Bigr\rvert_\text{spinrot}
  = \int \frac{d^dp_1}{(2\pi)^d} v_{1j}
  \frac{D\vec\sigma_1}{Dt}\Bigr\rvert_\text{spinrot}
\end{align}
can then be written using $\vec J_j^\perp = c_\perp
\alpha_\perp \magn \hat{\vec g}_j$ from Eq.~\eqref{eq:DJperp} as
\begin{align}
  \label{eq:6}
  \frac{D\vec J_j^\perp}{Dt}\Bigr\rvert_\text{spinrot}
  = \Omega_\text{mf}\, \hat{\vec e} \times \vec J_j^\perp.
\end{align}
The spin current precesses around the molecular field with frequency
\cite{mullin1992}
\begin{multline}
  \label{eq:spinrot}
  \Omega_\text{mf} = \frac{1}{\alpha_\perp \magn} \int
  \frac{d^dp_1}{(2\pi)^d} \, \frac{d^dp_2}{(2\pi)^d} \, 
  v_{1j} (v_{1j}-v_{2j}) (n_{1+}-n_{1-}) \\
  \times (n_{2+}-n_{2-}) \Re \mathcal T(\vec p_1+\vec p_2,\omega),
\end{multline}
which then determines the spin-rotation parameter
$\mu=-\Omega_\text{mf}\, \tau_\perp$.  For a momentum independent
interaction $\Re \mathcal T=2V_0$ this reduces to
$\Omega_\text{mf}=2V_0\magn/\hbar$ \cite{jeon1988}.

%%%%%%%%%%%%%%%%%%%%%%%%%%%%%%%%%%%%%%%%%%%%%%%%%%%%%%%%%%%%%%%%%%%%%%%%

\subsection{Longitudinal diffusion}
\label{sec:long}

For longitudinal spin diffusion one may linearize the distribution
matrix with a variation \eqref{eq:dsigmalong} that remains diagonal in
the spin indices.  Then also the linearized collision integral
\eqref{eq:matcoll} is diagonal, and following the standard derivation
one obtains the longitudinal scattering rate \cite{jeon1987,
  bruun2011, bruun2012, enss2012visc}
\begin{multline}
  \label{eq:taulonggen}
  \frac{1}{\tau_\parallel}
  = \frac{2\beta n}{(2\pi)^{2d}m^2n_+n_-}
  \int d^dq\, d^dk\, d\Omega\, k\frac{d\sigma}{d\Omega} \\
  \times n_{1+} n_{2-} \tilde n_{3+} \tilde n_{4-} k_j(k_j-k_j').
\end{multline}
The angular average yields
\begin{multline}
  \label{eq:taulongavg}
  \int d\Omega_{\vec q}\, d\Omega_{\vec k}\, d\Omega\,
  n_{1+} n_{2-} \tilde n_{3+} \tilde n_{4-} k_j(k_j-k_j') \\
  = \frac{S_d^3}{d} k^2 [I_{\ell=0}^2(a,b,c)-I_{\ell=1}^2(a,b,c)]
\end{multline}
in terms of the functions $I_\ell(a,b,c)$ defined in
Eqs.~\eqref{eq:Iell3d} and \eqref{eq:Iell2d}, and
\begin{multline}
  \label{eq:taulong}
  \frac{1}{\tau_\parallel}
  = \frac{2S_d^3\beta n}{d(2\pi)^{2d}m^2n_+n_-}
  \int_0^\infty dq\, q^{d-1} \int_0^\infty dk\, k^{d+2} \\
  \times \frac{d\sigma}{d\Omega}\, [I_0^2-I_1^2].
\end{multline}
In the Boltzmann limit $T\gg T_F$ one finds $I_{\ell=0}^2 \to z_+ z_-
\exp(-\beta\varepsilon_q/2) \exp(-2\beta\varepsilon_k)$ and
$I_{\ell=1}\to0$; hence \eqref{eq:taulong} converges toward the
transverse scattering rate \eqref{eq:tauperpboltz} independent of
polarization.  Likewise, in the unpolarized case $n/n_+n_- \to 4/n$
and $I_1\to0$, and the longitudinal scattering time converges toward
the transverse scattering time \eqref{eq:tauperpunpol} for all
temperatures.

%%%%%%%%%%%%%%%%%%%%%%%%%%%%%%%%%%%%%%%%%%%%%%%%%%%%%%%%%%%%%%%%%%%%%%%%

\section{Results}
\label{sec:results}

\subsection{Three dimensions}

\begin{figure}
  \centering
  \includegraphics[width=\linewidth]{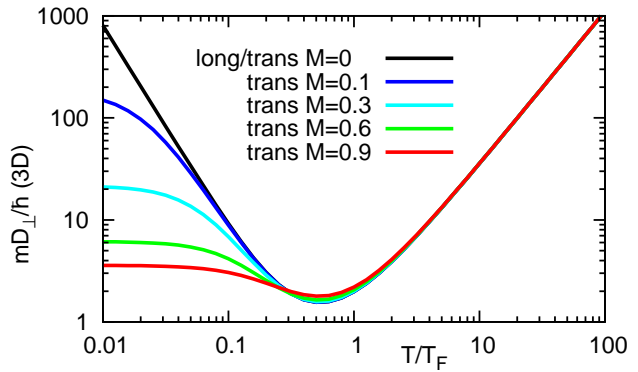}
  \caption{(Color online) Transverse and longitudinal spin
    diffusivities $D_\perp$ and $D_\parallel$ vs reduced temperature
    $T/T_F$ for different polarizations $M$ (top $M=0$ to bottom
    $M=0.9$) for the unitary Fermi gas in three dimensions.  The
    collision integral is computed using the vacuum $T$-matrix.}
  \label{fig:diff3d}
\end{figure}

Figure~\ref{fig:diff3d} shows the transverse and longitudinal spin
diffusivity $D_\perp$ and $D_\parallel$ vs reduced temperature $T/T_F$
in three dimensions.  Within kinetic theory the transverse and
longitudinal diffusivities are equal in two limits: for unpolarized
gases ($M=0$) at arbitrary temperature, and in the Boltzmann limit
$T\gg T_F$ for arbitrary polarization.  We therefore focus our study
on the polarized gas in the quantum degenerate regime where $D_\perp$
and $D_\parallel$ differ: as the polarization increases the transverse
diffusivity $D_\perp$ decreases at low temperatures and reaches a
finite value as $T\to0$.  This is in marked contrast to the
longitudinal diffusivity, which due to Pauli blocking diverges as
$D_\parallel \sim T^{-2}$ for a normal Fermi liquid.

\begin{figure}
  \centering
  \includegraphics[width=\linewidth]{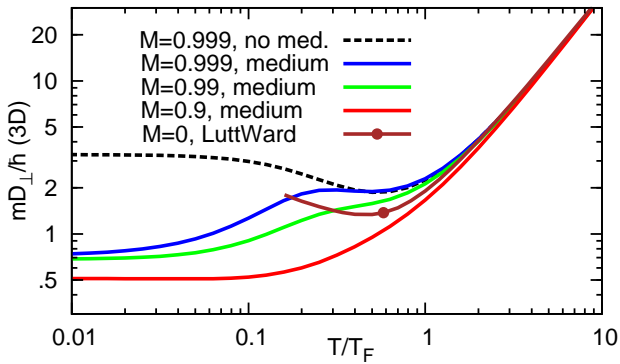}
  \caption{(Color online) Transverse spin diffusivity $D_\perp$ vs
    reduced temperature $T/T_F$ including medium scattering (solid
    lines: top $M=0.999$ to bottom $M=0.9$).  Due to increased
    scattering the medium diffusivity $D_\perp$ is lower than the
    vacuum diffusivity (dashed).  For comparison, the Luttinger-Ward
    curve (with circle) for the unpolarized gas \cite{enss2012spin}
    above $T_c \simeq 0.16\,T_F$ includes not only medium scattering
    but also the renormalization of the fermionic spectral function.}
  \label{fig:med3d}
\end{figure}

In Fig.~\ref{fig:diff3d} the diffusivities have been computed with the
vacuum scattering cross section, and the behavior agrees qualitatively
with that in the Born approximation \cite{jeon1989}.  However, as
explained in Sec.~\ref{sec:scatt}, in a systematic $1/N$ expansion
to leading order one has to use the \emph{medium} scattering cross
section in combination with the thermodynamic functions of the free
Fermi gas \cite{enss2012crit}.  The many-body $T$-matrix
\eqref{eq:tmat} has to be computed numerically with one integral;
hence the solution of the Boltzmann equation requires a
three-dimensional integral.  The resulting diffusivity $D_\perp$ is
shown in Fig.~\ref{fig:med3d}.  In the nondegenerate regime $T\gtrsim
T_F$ the effect of medium scattering is still small.  However, at
lower temperatures the medium strongly enhances scattering and leads
to a substantial suppression of the diffusivity, even more so away
from the fully polarized limit.  At the lowest temperatures $T\to0$ the
medium diffusivity still converges toward a finite value, as in the
vacuum scattering case.

For large polarization above the Clogston-Chandra\-sekhar limit
\cite{lobo2006}, the Fermi gas remains normal and the $T$-matrix is
well defined down to zero temperature.  For smaller polarization the
$T$-matrix develops a pole associated with the phase transition, and
the many-body $T$-matrix is reliable in the normal Fermi liquid phase
above the phase transition.  In the vicinity of the phase transition
kinetic theory becomes inaccurate, and one has to resort to more
elaborate transport calculations using, for instance, the
Luttinger-Ward framework based on the self-consistent $T$-matrix.  For
comparison, we plot the longitudinal spin diffusivity
$D_\parallel(M=0)$ (curve with circle) from a Luttinger-Ward
calculation \cite{enss2012spin}, which includes not only medium
scattering but also the renormalization of spectral functions on equal
footing, remaining regular down to $T_c \simeq 0.16\,T_F$.

\begin{figure}
  \centering
  \includegraphics[width=\linewidth]{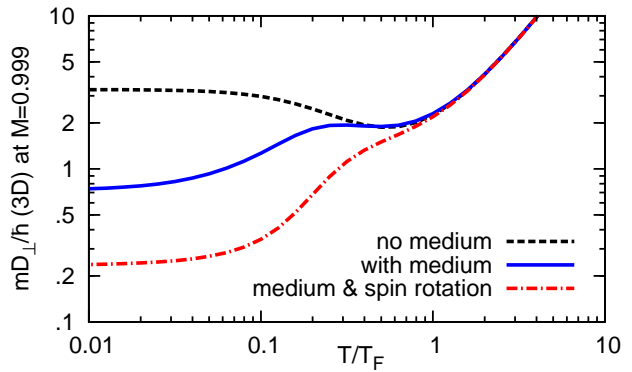}
  \caption{(Color online) Spin-rotation effect on the transverse spin
    diffusivity $D_\perp$ vs reduced temperature $T/T_F$ for large
    polarization $M=0.999$.  Dashed line without medium
    scattering, solid line with medium effects, and dash-dotted line
    including the spin-rotation effect Eq.~\eqref{eq:fulldiff} with
    spin-rotation parameter $\mu=-\Omega_\text{mf}\,\tau_\perp$, which
    further suppresses the diffusivity.}
  \label{fig:rot3d}
\end{figure}

Figure~\ref{fig:rot3d} shows the effect of spin rotation
\cite{leggett1970, mullin1992}: the spin current precesses around the
effective molecular field with frequency $\Omega_\text{mf}$, which
results in a lower transverse diffusivity $D_\perp$.  The molecular
field frequency \eqref{eq:spinrot} of the unitary Fermi gas in the
polaron limit $M\to1$ reaches $\Omega_\text{mf} \approx -1.2\,E_F$ for
$T=0$, which is twice the value of the chemical potential shift
\cite{schmidt2011}.  At large temperature, $\Omega_\text{mf}$ decays
as $T^{-2}$; hence $\mu = -\Omega_\text{mf}\, \tau_\perp \sim T^{-3/2}$
and there is no spin rotation in the Boltzmann limit.  Note that for
the 3D unitary Fermi gas the \emph{vacuum} $T$-matrix is purely
imaginary at $a^{-1}=0$ and leads to a vanishing molecular field;
$\Omega_\text{mf}$ is nonzero only for the \emph{medium} $T$-matrix,
which is used in Fig.~\ref{fig:rot3d}.  The full transverse spin
diffusivity $D_\perp$ (dash-dotted line) is strongly suppressed at low
temperatures but converges to the Boltzmann result at large
temperature where the molecular field vanishes.  This temperature
dependence provides an experimentally accessible signature of the
spin-rotation effect.  Note that the external magnetic field $\gamma
B$ does not affect the dynamics in the co-rotating frame
\cite{leggett1970}.

\begin{figure}
  \centering
  \includegraphics[width=\linewidth]{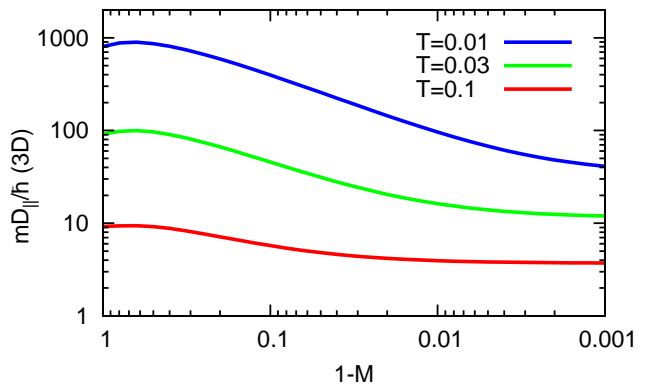}
  \caption{(Color online) Longitudinal spin diffusivity $D_\parallel$
    vs polarization $M$ at different temperatures $T/T_F$ (top
    $T/T_F=0.01$ to bottom $T/T_F=0.1$) for the 3D unitary Fermi gas
    (without medium scattering).}
  \label{fig:long3d}
\end{figure}

In Fig.~\ref{fig:long3d} the longitudinal spin diffusivity
$D_\parallel$ is plotted vs polarization.  At small polarization up to
about $50\%$ the diffusivity changes only slightly: it first increases
and then drops for larger polarization.  At very large polarization
above $98\%$, it eventually saturates to a finite value in the limit
$M\to1$.  This final value still depends on the temperature, roughly
as $D_\parallel \sim 0.37 (\hbar/m)\, (T/T_F)^{-1}$.

\subsection{Two dimensions}

\begin{figure}
  \centering
  \includegraphics[width=\linewidth]{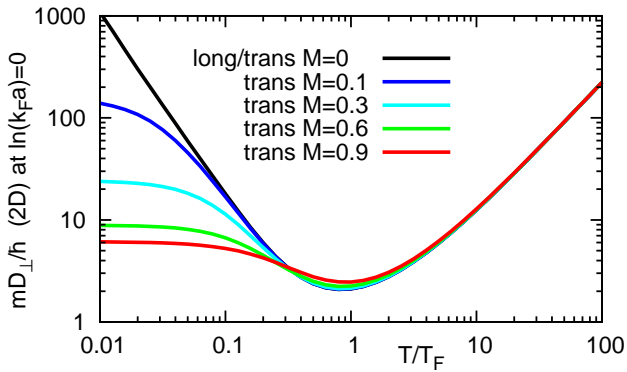}  
  \caption{(Color online) Transverse spin diffusivity $D_\perp$ vs
    reduced temperature $T/T_F$ for different polarizations $M$ (top
    $M=0$ to bottom $M=0.9$) for a strongly interacting 2D Fermi gas
    with interaction parameter $\ln(k_Fa_{2D})=0$ (without medium
    scattering).}
  \label{fig:diff2d}
\end{figure}

The spin diffusivity in 2D has recently attracted
interest after spin-echo measurements in a transversely polarized spin
state in an ultracold gas of fermionic atoms \cite{koschorreck2013}.
The decay of magnetization over time allows one to infer the spin
diffusivity, and very low values for $D_\perp$ have been found in the
strongly interacting regime.  In order to understand these results, we
first compute the transverse and longitudinal spin diffusivities in 2D
without medium scattering and find that they exhibit a qualitatively
similar behavior as in the 3D case, as shown in Fig.~\ref{fig:diff2d}.

\begin{figure}
  \centering
  \includegraphics[width=\linewidth]{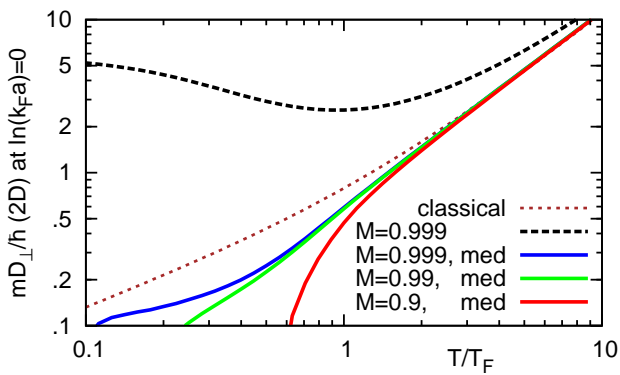}  
  \caption{(Color online) Transverse spin diffusivity $D_\perp$ vs
    reduced temperature $T/T_F$ in 2D including medium scattering at
    strong interaction $\ln(k_Fa_{2D})=0$ (solid lines: top $M=0.999$
    to bottom $M=0.9$).  The dashed line is for vacuum scattering,
    while the dotted curve illustrates the classical result
    \eqref{eq:Dperp2D} in the Boltzmann limit.}
  \label{fig:med2d}
\end{figure}

However, the effect of medium scattering is even more pronounced in 2D
than in 3D and can suppress the diffusivity by more than 1 order of
magnitude at low temperature (see Fig.~\ref{fig:med2d}).  For very
large polarization $M=0.999$ the diffusivity appears to saturate
around $T/T_F=0.1$ near $D_\perp \approx 5\,\hbar/m$ without medium
scattering, and near $D_\perp \approx 0.1\,\hbar/m$ if the medium is
included in the calculation.  While Pauli blocking alone increases
$D_\perp$ (dashed curve), the medium compensates this effect and leads
to values of $D_\perp$ closer to the classical result
\eqref{eq:Dperp2D} (dotted curve).  The suppression of the diffusivity
for smaller polarization signals the appearance of a superfluid
density at low temperature, which would lead to a pole in the
non-selfconsistent $T$-matrix and a diverging collision integral
\cite{enss2012visc}.

\begin{figure}
  \centering
  \includegraphics[width=\linewidth]{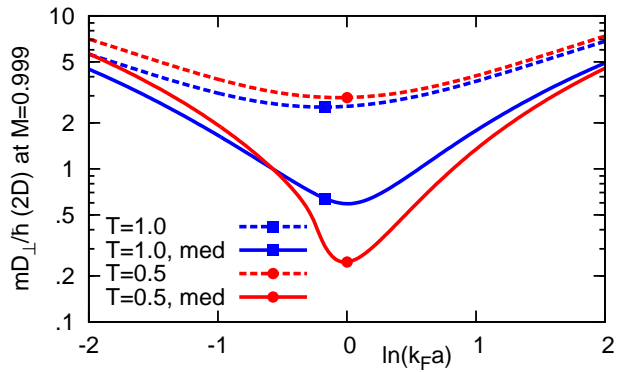}
  \caption{(Color online) Transverse spin diffusivity $D_\perp$ vs
    interaction strength $\ln(k_Fa_{2D})$ at fixed polarization
    $M=0.999$ and temperatures $T/T_F=1$ (blue/square),
    $T/T_F=0.5$ (red/circle).  The dashed lines denote the
    diffusivity without medium effects, while the solid lines include
    medium scattering.}
  \label{fig:int2d}
\end{figure}

The interaction dependence of the transverse diffusivity is shown in
Fig.~\ref{fig:int2d} for two values of the temperature in the quantum
degenerate regime.  At fixed polarization $M=0.999$, the suppression
by medium effects (solid vs dashed lines) is most pronounced in the
strongly interacting region $-1 \lesssim \ln(k_Fa_{2D}) \lesssim 1$,
while at weak coupling the medium effects lower the diffusivity only
slightly.  The values of $D_\perp$ in Fig.~\ref{fig:int2d} come close
to $D_\perp = 0.25(3)\,\hbar/m$, measured in a recent 2D spin-echo
experiment \cite{koschorreck2013}, although the measured minimum
around $\ln(k_Fa_{2D})=0$ is more shallow than in our calculation.

In order to make a detailed comparison of our transport calculation
for the homogeneous system with experiments in a trap geometry, it
would be useful to measure the diffusivity for evolution times
shorter than the trap period in order to minimize the effects of the
trap.  Measuring the temperature dependence of the diffusivity would
also provide a much more sensitive comparison of theory and
experiment, in particular regarding the spin-rotation effect displayed
in Fig.~\ref{fig:rot3d}.

%%%%%%%%%%%%%%%%%%%%%%%%%%%%%%%%%%%%%%%%%%%%%%%%%%%%%%%%%%%%%%%%%%%%%%%%

\section{Conclusion}
\label{sec:concl}

We have presented a kinetic theory for transverse and longitudinal
spin diffusion in strongly interacting Fermi gases in two and three
dimensions based on the many-body $T$-matrix.  We find a significant
suppression of the spin diffusivities at low temperatures and strong
coupling due to medium scattering beyond the Born approximation.  The
results are consistent with the very low transverse spin diffusivity
$D_\perp$ observed in a recent 2D spin-echo
experiment \cite{koschorreck2013} at strong interaction.  Our analysis
includes the Leggett-Rice effect of spin rotation by a molecular field
\cite{leggett1970}, which further lowers the transverse diffusion
coefficient of a polarized gas.  It will be interesting to study the
role of mean-field corrections to the quasiparticle dispersion
relation \cite{chiacchiera2009} in a future work.

For small polarization below the Clogston-Chandra\-sekhar limit, the
interacting Fermi gas exhibits a phase transition toward
superfluidity and the ladder approximation for the $T$-matrix may
have to be amended by particle-hole fluctuations near the transition.
In this case it would be worthwhile to compute transverse spin
transport also using other theoretical approaches which go beyond a
quasiparticle description, such as the Luttinger-Ward
\cite{enss2012spin} or Monte Carlo methods \cite{wlazlowski2013}, but
we expect that the qualitative features will be similar.

I am grateful to Michael K\"ohl, Richard Schmidt, and Joseph
H. Thywissen, and Wilhelm Zwerger for fruitful discussions.

%%%%%%%%%%%%%%%%%%%%%%%%%%%%%%%%%%%%%%%%%%%%%%%%%%%%%%%%%%%%%%%%%%%%%%%%

\end{document}